\documentclass{elsart}

\usepackage{graphicx}

\usepackage{amssymb}

\begin{document}

\begin{frontmatter}



\title{N-qubit entanglement via the $J_y^2$-type collective interaction}


\author[GAT,ITP]{D.L. Zhou\corauthref{cor}},
\corauth[cor]{Corresponding author.} \ead{dz34@mail.gatech.edu}
\author[MIT]{B. Zeng},
\ead{zengbei@MIT.EDU}
\author[TSC]{J.S. Tang},
\ead{tjs98@mail.tsinghua.edu.cn}
\author[TSP]{Z. Xu},
\ead{zx-dmp@mail.tsinghua.edu.cn}
\author[GAT,ITP]{L. You}
\ead{ly14@mail.gatech.edu}

\address[GAT]{School of
Physics, Georgia Institute of Technology, Atlanta, GA 30332, USA}

\address[ITP]{Institute of Theoretical Physics, The Chinese Academy
of Sciences, Beijing, 100080, China}

\address[MIT]{Department of Physics, Massachusetts Institute of
Technology, Cambridge, MA 02139, USA}

\address[TSC]{Department of Computer Science and Technology,
Tsinghua University, Beijing, 100084, China}

\address[TSP]{Department of Physics, Tsinghua University, Beijing,
100084, China}

\begin{abstract}
% Text of abstract
We investigate quantum correlations of the $N$-qubit states via a
collective pseudo-spin interaction ($\propto J_y^2$) on arbitrary
pure separable states for a given interval of time. Based on this
dynamical generation of the $N$-qubit maximal entangled states, a
quantum secret sharing protocol with $N$ continuous classical
secrets is developed.
\end{abstract}


\end{frontmatter}

% main text
%\section{}
%\label{}
\section{Introduction}
After more than ten years of active research and development in
quantum information and quantum computation, quantum information
science has become a major theme of contemporary physics research
\cite{Ni}. The understanding and characterization of quantum
entanglement has emerged as a key fundamental issue. Despite much
intense research efforts \cite{Dur,Coffman,Bennett}, we are still
far from forming a complete picture of multi-partite quantum
correlations, especially in the limit of (large) N-qubit systems.
In recent years, however, several studies have provided much
insight into the controlled generation of a special class of
N-qubit entangled states, the so-called maximally entangled, or
N-GHZ states \cite{Green,ut,pan}.

M\"olmer and  S\"orenson \cite{Molmer} first discovered that N-GHZ
states can be created with a $J_y^2$-type collective interaction
\cite{jaksch,Chat}. Their protocol as engineered in a linear ion
trap, has led to the experimental demonstration of the largest
Schrodinger cat state, an N-GHZ state of four trapped ions
\cite{Sacket}. It was soon realized that the same interaction
naturally occurs in a two component atomic condensate and can be
utilized for the deterministic generation of entanglement atomic
condensates \cite{You}.

Two somewhat puzzling points from these recent results stimulated
the current study; 1) To generate an N-GHZ state with the
$J_y^2$-type interaction, the initial states were all product
states of the same single qubit pure state, which is itself an
eigenstate of $J_{\vec x_\perp}$, i.e. the collective spin
component along a direction orthogonal to $\hat y$. 2) The N-GHZ
protocol of $J_y^2$ shows an even/odd qubit number parity; it
requires an extra single particle interaction term depending on
whether $N$ is even or odd, (see, for instance Ref.
\cite{Molmer}).

This article provides an analytic approach to study the target
maximally entangled N-GHZ states from the $J_y^2$-type interaction
when operated on any initial separable pure state. In addition to
presenting a clear resolution of the above two mentioned puzzles
from earlier studies of N-GHZ state creation, our result provides
a general and convenient frame to discuss N-qubit entanglement.

This paper is organized as follows; First, we provide a general
proof of a proposition using techniques developed for quantum
error-correcting codes and cluster states \cite{Got,Bri,Rau1}. We
show that a $J_y^2$-type interaction generates N-GHZ states for
all N-qubit separable real pure states. Based on this extension,
we provide a secret sharing protocol in which no single party can
obtain his secret key without cooperations. We then suggest an
optimal basis for projecting an arbitrary qubit onto the
orthogonal plane to find the largest N-GHZ component as created by
the $J_y^2$-type interaction from an arbitrary N-qubit separable
pure state; along the operation steps for the projection, we find
that the system dynamics can be properly accounted for in terms of
a set of complete and orthogonal N-GHZ states. We finally conclude
with some commentary remarks.

\section{Generating a N-GHZ state with a $J_y^2$-type interaction
from an unknown separable real state}

\begin{figure}[htbp]
\begin{center}
\includegraphics[width=2.in]{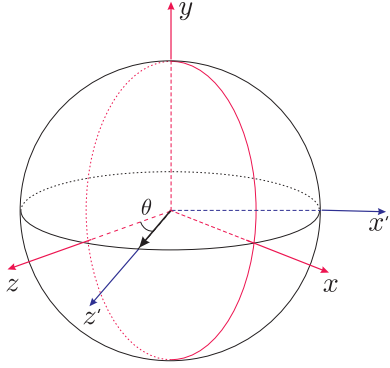}
\end{center}
\caption{An arbitrary pure qubit state in the equator $z$-$x$
plane of the north-south axis $\hat y$ (of $\vec J_y$), and the
corresponding locally rotated coordinate $(x',y,z')$.}
\label{fig1}
\end{figure}

Using the geometrical Bloch sphere representation of a qubit, an
arbitrary pure state is represented by the vector from origin to
the point $(\theta,\phi)$ on the sphere. The special class of real
separable pure states
\begin{eqnarray}
|\psi\rangle=\prod_{k=1}^{N}\left(\cos{\frac {\theta_k}
2}|0\rangle^{(k)} +\sin{\frac {\theta_k} 2}|1\rangle^{(k)}\right),
\label{esp}
\end{eqnarray}
corresponds to all N-qubit Bloch vectors lie in the $z$-$x$ plane
(as in Fig. \ref{fig1} for a real qubit), with $N$ unknown real
parameters $\{\theta_i\}$ ${i=1,\cdots,N}$.

To generate a maximally entangled N-GHZ state from the sparable
pure state (\ref{esp}), the most straightforward approach would be
to invoke the unitary transformation
\begin{eqnarray}
U_N=\frac {1} {\sqrt{2}}
\left(I+i\prod_{k=1}^{N}{\sigma_{y}^{(k)}}\right).
\end{eqnarray}
This is difficult to realize for N-qubit physical systems,
however, for it needs $N$-particle interactions. The $J_y^2$-type
interaction, on the other hand, involves only binary two-body
interactions, and can be engineered in several known physical
systems \cite{Molmer,Sacket,You}. We now prove the following {\bf
proposition}: The unitary operator
\begin{eqnarray}
S=\prod\limits_{i,j=1,i<j}^{N}S_{ij}, \label{uo}
\end{eqnarray}
with
\begin{eqnarray}
S_{ij}=(I+\sigma_{y}^{(i)}+\sigma_{y}^{(j)}-\sigma_{y}^{(i)}\sigma_{y}^{(j)})/2,
\end{eqnarray}
transforms a pure separable N-qubit real state $|\psi\rangle$
(\ref{esp}) into a maximally entangled N-GHZ state
$|\psi_M\rangle$, i.e. $|\psi_M\rangle=S|\psi\rangle$.

Before proving our proposition, we note the following structure of
a qubit state. As shown in Fig. \ref{fig1}, we define new spin
matrices with respect to the locally rotated axis
\begin{eqnarray}
\sigma_{z^{\prime}}^{(i)}&=&
\sigma_z^{(i)}\cos{\theta_i}+\sigma_x^{(i)}\sin{\theta_i},\\
\sigma_{x^{\prime}}^{(i)}&=&
-\sigma_z^{(i)}\sin{\theta_i}+\sigma_x^{(i)}\cos{\theta_i}.
\end{eqnarray}
Clearly, the eigenstates of the $\sigma_{z^{\prime}}^{(i)}$ with
eigenvalues $1$ and $-1$ are given by
\begin{eqnarray}
|0\rangle_{z^{\prime}}^{(i)}&=&\cos \frac {\theta_i} {2}
|0\rangle^{(i)}
+ \sin \frac {\theta_i} {2} |1\rangle^{(i)},\\
|1\rangle_{z^{\prime}}^{(i)}&=&\sigma_{x^{\prime}}^{(i)}|0\rangle_{z^{\prime}}^{(i)}
=\cos \frac {\theta_i} {2} |1\rangle^{(i)} - \sin \frac {\theta_i}
{2} |0\rangle^{(i)}.
\end{eqnarray}
Similarly, the eigenstates of $\sigma_{x^{\prime}}^{(i)}$ are
\begin{eqnarray}
|0\rangle_{x^{\prime}}^{(i)}&=&\frac 1
{\sqrt{2}}(|0\rangle_{z^{\prime}}^{(i)}+|1\rangle_{z^{\prime}}^{(i)})=
\cos \frac {\theta_i+\frac {\pi} {2}} {2} |0\rangle^{(i)}
+ \sin \frac {\theta_i+\frac {\pi} {2}} {2} |1\rangle^{(i)}, \label{bs9}\\
|1\rangle_{x^{\prime}}^{(i)}&=&\sigma_{y}^{(i)}|0\rangle_{x^{\prime}}^{(i)}
=-i\sin \frac {\theta_i+\frac {\pi} {2}} {2}|0\rangle^{(i)}+i\cos
\frac {\theta_i+\frac {\pi} {2}} {2} |1\rangle^{(i)}. \label{bs10}
\end{eqnarray}

It is easy to check that
$\{{\sigma}_{x^{\prime}}^{(i)},{\sigma}_y^{(i)}
,{\sigma}_{z^{\prime}}^{(i)}\}$ satisfy all the commutation
relations of Pauli operator, i.e. they form a group of Pauli
matrices of a spin $1/2$ operator.

With the above redefinition
($|\psi\rangle=\sigma_{z^{\prime}}^{(i)}|\psi\rangle$), our
proposition is equivalent to $|\psi_M\rangle=S\prod_{i=1}^N
|0\rangle_{z^{\prime}}^{(i)}$, or
\begin{eqnarray}
|\psi_M\rangle=S\sigma_{z^{\prime}}^{(i)}S^{\dag}|\psi_M\rangle.
\end{eqnarray}
An easy calculation leads to
\begin{eqnarray}
S_{ij}\sigma_{z^{\prime}}^{(i)}S_{ij}^{\dag}&=&
\sigma_{z^{\prime}}^{(i)}\otimes\sigma_{y}^{(j)},\nonumber\\
S_{ij}\sigma_{z^{\prime}}^{(j)}S_{ij}^{\dag}&=&\sigma_{y}^{(i)}
\otimes\sigma_{z^{\prime}}^{(j)}, \label{eqproof}
\end{eqnarray}
and
\begin{eqnarray}
S_{ij}\sigma_{z^{\prime}}^{(k)}S_{ij}^{\dag}&=&
\sigma_{z^{\prime}}^{(k)},\hskip 12pt \;k\neq i,j,\nonumber\\
S_{ij}\sigma_{y}^{(k)}S_{ij}^{\dag}&=&\sigma_{y}^{(k)},\hskip 12pt
\; \forall\,k=1,\cdots,N.
\end{eqnarray}
Thus we get
\begin{eqnarray}
S\sigma_{z^{\prime}}^{(i)}S^{\dag}=\sigma_{z^{\prime}}^{(i)}
\prod_{j=1,j\neq i}^N\sigma_y^{(j)},
\end{eqnarray}
i.e.
\begin{eqnarray}
|\psi_M\rangle=\sigma_{z^{\prime}}^{(i)}\prod_{j=1,j\neq
i}^N\sigma_y^{(j)}|\psi_M\rangle.
\end{eqnarray}

The above $N$ equations uniquely determines the state
$|\psi_M\rangle$. We note that for every different pair
$\{i,i'\}$,
\begin{equation}
\left(\sigma_{z^{\prime}}^{(i)}\prod_{j\neq i
}\sigma_y^{(j)}\right) \left(
\sigma_{z^{\prime}}^{(i')}\prod_{j'\neq i'
}\sigma_y^{(j')}\right)=\sigma_{x^{\prime}}^{(i)}\otimes\sigma_{x^{\prime}}^{(i')}.
\end{equation}
Thus, the state $|\Psi_M\rangle$ is limited to a two dimensional
Hilbert space and can be expanded according to
\begin{equation}
|\psi_M\rangle=\alpha\prod_{i=1}^{N}|0\rangle_{x^{\prime}}^{(i)}
+\beta\prod_{i=1}^{N}|{1}\rangle_{x^{\prime}}^{(i)}.
\end{equation}
The coefficients $\alpha$ and $\beta$ can be determined from
\begin{equation}
\sigma_{z^{\prime}}^{(1)}\prod_{j=2}^{N}\sigma_y^{(j)}|\psi_M\rangle=|\psi_M\rangle,
\end{equation}
and this leaves us with the explicit expression for
$|\psi_M\rangle$,
\begin{eqnarray}
|\psi_M\rangle=\frac{1}{\sqrt{2}}\left(\prod_{i=1}^{N}|0\rangle_{x^{\prime}}^{(i)}
+i\prod_{i=1}^{N}|1\rangle_{x^{\prime}}^{(i)}\right), \label{mes}
\end{eqnarray}
which is a maximally entangled state. This completes our proof.

We now construct an equivalent Hamiltonian corresponding to the
unitary operation Eq. (\ref{uo}). We note that
$$S_{jk}=\exp{[-i{\pi}/{4}+i{\pi}{\sigma}_y^{(j)}/{{4}}+i{\pi}{\sigma}_y^{(k)}/{{4}}
-i{\pi}{\sigma}_y^{(j)}{\sigma}_y^{(k)}/{4}]},$$ which leads to
\begin{eqnarray}
S=&&\exp{\left(-iN(N-1)\frac{\pi}{4}\right)}
\exp{\left(i(N-1)\frac{\pi}{{4}}\sum_{j=1}^N{\sigma}_y^{(j)}\right)}\nonumber\\
&&\times\exp{\left(-i\frac{\pi}{4}\sum_{j,k=1,j<k}^{N}
{\sigma}_y^{(j)}{\sigma}_y^{(k)}\right)},
\end{eqnarray}
i.e. the unitary time evolution from an interaction Hamiltonian
\begin{eqnarray}
H_I\propto\sum\limits_{i,j=1,i<j}^{N}
{\sigma}_y^{(i)}{\sigma}_y^{(j)}\sim J_y^2,
\end{eqnarray}
where $J_y=\sum\limits_{i=1}^{N} {\sigma}_y^{(i)}/2$ is the $y$
component of the total spin of the system and $\sim$ means apart
from a constant. It is well-known that the interaction Hamiltonian
$uJ_y^2$ can be realized in many physical systems, such as trapped
ions \cite{Molmer} and Bose-Einstein condensed atoms \cite{You}.

\section{A quantum secret sharing protocol}

Before extending the above discussion to arbitrary pure separable
initial state, we want to comment on the maximally correlated
nature of state (\ref{mes}) despite of the unknown $\theta_i$'s. A
simple criticism could naively point to the maximally entangled
state (\ref{mes}) as a mathematical formula because its basis
states [Eqs. (\ref{bs9}) and (\ref{bs10})] are unknown. If
$\theta_i$ were known, the initial state (\ref{esp}) becomes
essentially the same as in the previous works \cite{Molmer,You},
apart from local unitary transformations (basis rotations with
known angles $\theta_i$ for every qubits). So what is new in this
work? To address this important question, we have developed a
quantum secret sharing protocol \cite{Hil,Cle,Ter} to reveal the
powerful multi-party quantum correlations of the state (\ref{mes})
irrespective of whether the basis states of the constituent qubits
are known or not. The naive criticism arises due to a general lack
of adequate understanding for multi-party entanglement, or more
explicitly the lack of a reasonable multi-party entanglement
measure. If there were such a measure, the entanglement of
(\ref{mes}) is obviously unaffected by the $\theta_i$'s.

Our quantum secret sharing protocol involves a queue ($\theta_1$,
$\theta_2$,$\cdots$, $\theta_N$). The unknown angles $\theta_i$'s
can be encoded into the N-qubit initial state (\ref{esp}), the
unitary operation (\ref{uo}) is then affected to create the
maximally entangled state (\ref{mes}). Giving the $j$-th qubit to
the $j$-th party, we accomplish a quantum secret sharing whereby
no individual parties can obtain any information about the queue
alone, but the N parties can cooperate to uncover the complete
queue. In fact, in this simple protocol, each individual party
cannot even obtain its own key $\theta_j$ alone, because the
reduced state for the $j$-th qubit is completely mixed
\begin{eqnarray}
\rho_i^{(1)}=\textrm
{Tr}_{1,2,\cdots,j-1,j+1,\cdots,N}(|\psi_M\rangle\!\langle
\psi_M|)=\frac {1} {2}I.
\end{eqnarray}
To uncover the queue cooperatively, the N parties can simply
execute the inverse of the operation (\ref{uo}). Each of the
parties can then find its $\theta_j$ by local measurements in the
initial real separable pure state (\ref{esp}).

This protocol shows that the maximally entangled state (\ref{mes})
remains useful due to the strong underlying N-qubit correlations
despite of the unknown parameters $\theta_j$'s. Our present work
is therefore different from earlier ones \cite{Molmer,You} where
only initial states with known $\theta_i\, (\theta_i\equiv 0)$
were discussed. Incidently, we note that the above protocol for
quantum secret sharing is not a threshold scheme \cite{Cle}
because it turns out that any $m$ ($1<m<N$) parties can work
together to determine their own keys up to two possible choices
without needing the other $N-m$ qubits. For example, in the case
of $m=2$, the reduced density matrix for two parties $i$ and $j$
becomes
\begin{eqnarray}
\rho^{(ij)}&=&\frac{1}{2}|0\rangle_{x^{\prime}}^{(i)}\mbox{}^{(i)}_{x^{\prime}}\!\langle
0|\otimes|0\rangle_{x^{\prime}}^{(j)}\mbox{}^{(j)}_{x^{\prime}}\!\langle
0|+\frac{1}{2}|1\rangle_{x^{\prime}}^{(i)}\mbox{}^{(i)}_{x^{\prime}}\!\langle
1|\otimes|1\rangle_{x^{\prime}}^{(j)}\mbox{}^{(j)}_{x^{\prime}}\!\langle
1|,
\end{eqnarray}
which can be completely determined by measurements performed by
parties $i$ and $j$. It is worth emphasizing that the above
decomposition of the reduced density matrix is unique since it is
not only an eigen-decomposition but also a decomposition with
separable state components. However, we cannot distinguish the
component
$|0\rangle_{x^{\prime}}^{(i)}|0\rangle_{x^{\prime}}^{(j)}$ from
$|1\rangle_{x^{\prime}}^{(i)}|1\rangle_{x^{\prime}}^{(j)}$, thus
we have two possible choices for the parameter pair ($\theta_i$,
$\theta_j$). This discussion for $m=2$ remains valid for cases of
$2<m<N$, \textit{i.e.} $m$ parties can determine their secret
keys $\theta$s upto two possible alternative choices.

Before investigating the entangling dynamics of the $J_y^2$-type
interaction for more general N-qubit initial states, we hope to
contrast our secret sharing protocol with several previously known
schemes \cite{Hil,Cle,Ter}. An obvious difference for our scheme
is the need of many copies of identical quantum states. Previous
protocols, on the other hand, usually require only a single copy
of quantum state. However, the secret being shared by the
multi-party through our protocol are unknown continuous classical
variables, rather than basis states $|0\rangle$ or $|1\rangle$ of
each qubit. To faithfully reconstruct a continuous variable from a
qubit state always requires many identical copies. Furthermore,
despite of a large number of copies of quantum states as required
for our protocol, nothing about the shared secret can be obtained
locally.  This thus constitutes a legitimate secret sharing
protocol based on the quantum correlations of the N-GHZ type
states with unknown parameters.

\section{N-qubit entanglement with a $J_y^2$-type interaction
from an arbitrary pure separable state}

\begin{figure}[htbp]
\begin{center}
\includegraphics[width=2.in]{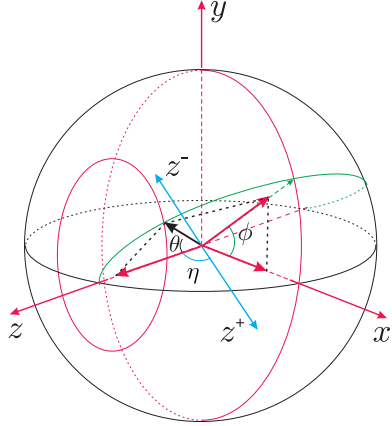}
\end{center}
\caption{An arbitrary qubit state can be decomposed into
projections along orthogonal states in the equator $z$-$x$ plane.}
\label{fig2}
\end{figure}

The proposition established earlier enables a clear picture for
the generation of N-GHZ states from an arbitrary separable initial
pure state using an interaction $\sim J_y^2$. We now consider the
more general initial separable pure state of N qubits
\begin{eqnarray}
|\psi\rangle=\prod_{k=1}^{N}|\psi\rangle^{(k)},\label{genstat}
\end{eqnarray}
with
\begin{equation}
|\psi\rangle^{(k)}=\cos{\frac {\theta_k} 2}|0\rangle^{(k)}
+\sin{\frac {\theta_k} 2}e^{i\phi_k}|1\rangle^{(k)},
\end{equation}
where all single qubit states are now specified by
$(\theta_j,\phi_j)$ as illustrated in Fig. \ref{fig2}. Clearly,
the arbitrary wavefunction $|\psi\rangle$ can be expanded into the
product basis $|0\rangle_z^{(i)}$ and $|1\rangle_z^{(i)}$ of the
N-qubits. Since each of the product basis state, e.g. a state
$\underbrace{|0\rangle_z^{(i_1)}\cdots|0\rangle_z^{(i_m)}}_{m}
\underbrace{|1\rangle_z^{(j_1)}\cdots|1\rangle_z^{(j_{N-m})}}_{N-m}$
containing $m$-0s and $(N-m)$-1s, evolves into its corresponding
maximally entangled N-GHZ state, this expansion enables a description
of the N-qubit wavefunction in terms of a complete
 orthonormal N-GHZ basis.

To characterize the entanglement properties of the above N-qubit
state, it is desirable to find the maximum coefficient into one
arbitrary N-GHZ state \cite{Zeng}. The simple geometrical Bloch
sphere representation might naively lead us to a decomposition of
the state $(\theta_j,\phi_j)$, into a component in the orthogonal
$z$-$x$ plane, plus a parallel component along the $y$-axis.
Unfortunately, this is incorrect because points on the Bloch
sphere are spinors, not real vectors. Two points on the opposite
side of the sphere, correspond to two vectors, not in
(anti-)parallel, but in fact orthogonal. Perhaps somewhat strange,
any state $(\theta_j,\phi_j)$ can be decomposed into two
orthogonal components along the opposite directions of a line
through the origin.

We now study the optimal decomposition of state (\ref{genstat})
with orthogonal states in the $z$-$x$ plane (along the $z^+$ and
$z^-$ axis). We assume the optimal orthogonal states for $k$-th
qubit are
\begin{eqnarray}
|\phi^{+}\rangle^{(k)}&=&\cos{\frac {\eta_k} 2}|0\rangle^{(k)}
+\sin{\frac
{\eta_k} 2}|1\rangle^{(k)},\\
|\phi^{-}\rangle^{(k)}&=&-\sin{\frac {\eta_k} 2}|0\rangle^{(k)}
+\cos{\frac {\eta_k} 2}|1\rangle^{(k)}.
\end{eqnarray}
The initial state can then be expanded according to
\begin{eqnarray}
|\psi\rangle&=&\prod_{k=1}^{N}\sum_{s_k=+,-}
{^{(k)}\langle \phi^{s_k}|\psi\rangle^{(k)}}|\phi^{s_k}\rangle^{(k)}\nonumber\\
&=&\sum_{\{s_k\}} \left(\prod_{k=1}^{N}{^{(k)}\langle
\phi^{s_k}|\psi\rangle^{(k)}}\right)\prod_{k=1}^{N}|\phi^{s_k}\rangle^{(k)}.
\end{eqnarray}
Following this decomposition, the final state after the evolution
by the $J_y^2$-type interaction becomes
\begin{eqnarray}
|\psi_F\rangle=\sum_{\{s_k\}}\left(\prod_{k=1}^{N}{_k\langle
\phi^{s_k}|\psi\rangle^{(k)}}\right)
S\prod_{k=1}^{N}|\phi^{s_k}\rangle^{(k)}. \label{eal}
\end{eqnarray}
According to our proposition,
$S\prod_{k=1}^{N}|\phi^{s_k}\rangle^{(k)}$ constitutes a maximally
entangled orthonormal basis. Thus, equation (\ref{eal}) provides
an alternative superposition of the $2^n$ orthogonal maximum
entanglement states.

\begin{figure}[htbp]
\begin{center}
\includegraphics[width=3.25in]{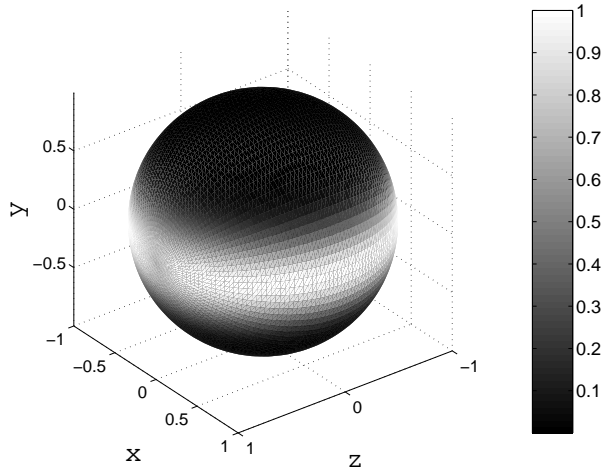}
\end{center}
\caption{The maximum N-GHZ state probability distribution for a
condensate (of $N=30$ atoms) in an initial state $(\theta,\phi)$
on the Bloch sphere.} \label{fig3}
\end{figure}

The optimal decomposition is found by maximizing the probability
of one particular N-GHZ state basis state,
\begin{eqnarray}
p(\eta_k)&\equiv&\left|{^{(k)}\langle \phi^{+}|\psi\rangle^{(k)}}\right|^2\nonumber\\
&=&(1+\cos\theta_k\cos\eta_k+\sin\theta_k
\cos\phi_k\sin\eta_k)/{2}, \hskip 24pt
\end{eqnarray}
where $\eta_k\in(-\pi, \pi]$. This maximum is then specified by
\begin{eqnarray}
\frac {\partial p(\eta_k)} {\partial \eta_k}&=&0,\\
\frac {\partial^2 p(\eta_k)} {\partial {\eta_k}^2}&<&0,
\end{eqnarray}
which gives
\begin{eqnarray}
&&\tan \eta_k=\tan\theta_k \cos \phi_k,\label{angle}\\
&&\sin \eta_k \cos \phi_k >0.
\end{eqnarray}
These conditions determine $\eta_k$ completely. The solution
actually has a simple geometrical interpretation: the projection
of $\vec{n}=(\sin\theta\cos\phi,\sin\theta\sin\phi,\cos\theta)$
onto the $z$-$x$ plane is $(\cos \theta,\sin \theta \cos \phi)$,
which gives the angle defined by equation (\ref{angle}). The
maximal probability becomes the product of the following for every qubit
\begin{eqnarray}
p^{\rm max}(\eta_k)={1\over 2}\left(1+\sqrt{1-\sin^2
\theta_k\sin^2\phi_k\,}\right).
\end{eqnarray}
In order to make a conclusive statement for NPT entanglement of
the N-qubits, $\prod_{k=1}^N p^{\rm max}(\eta_k)$ has to be larger
than $1/2$ \cite{Zeng}. As a special case, we consider Bose
condensed atoms in the same single qubit state
\begin{equation}
|\psi\rangle=\left(\cos \frac {\theta} {2} |0\rangle +\sin \frac
{\theta} {2} e^{i\phi}|1\rangle\right)^N.
\end{equation}
The maximum probability then becomes
$(1+\sqrt{1-\sin^2\theta\sin^2\phi\,})^N/2^N$, which is
illustrated on the Bloch sphere as in Fig. \ref{fig3}. We find
that with increasing atom numbers, states contain large N-GHZ
components become increasingly localized to near the $z$-$x$
equator plane. When $\phi=0$, i.e. for a real qubit state, we
always obtain a pure N-GHZ state. For $\phi\neq 0$, the optimal
projection is such that $\theta$ should be as close as possible to
the north/south pole, and the final state will be increasingly
farther away from the N-GHZ state with increasing $N$. Although,
as was shown earlier \cite{You}, one can simply apply a Raman
coupling to rotate the collective interaction $J_y^2$ into the
appropriate orthogonal direction to obtain a maximally entangled
state.

\section{Discussion and conclusion}

For practical applications, our result relives the restrictions on
the N-qubit initial state to an arbitrary separate real pure
state. It also relieves the odd/even-N constraint associated with
N-qubit maximum entanglement using the collective $J_y^2$-type
interaction, i.e. one can simply affect the unitary evolution Eq.
(\ref{uo}), or $S_{ij}$ among all pairs irrespective whether the
total number of qubit N being even or odd. Furthermore, when
applied to ensembles of cold atoms, it now only requires the two
state atoms interacting with each other with the same fixed
collision strength, the atoms do not in principle have to be Bose
condensed into the same spatial (condensate) mode. For that
matter, they do not even have to be condensed. Although condensed
atoms are believed to less susceptible to decoherence.

In conclusion, we have provided an analytic approach for
describing the controlled generation of N-GHZ states using a
collective interaction $\sim J_y^2$. We have derived the maximum
N-GHZ projection of the target state evolved from an arbitrary
separable initial pure state and provided a simple procedure to
characterize quantum correlations of the target state. We have
developed a quantum secret sharing protocol based on the general
N-GHZ states (\ref{mes}), thus illuminated their multi-party
correlations despite of the unknown parameters $\theta_j$'s. This
is a rare example from an initial pure product state, whose
generalized GHZ type quantum correlation due to the dynamic
evolution with $\sim J_y^2$ can be efficiently characterized and
understood at selected times. We believe our work will stimulate
new experimental efforts to generate maximally entangled N-GHZ
states.

We thank Dr. Peng Zhang for insightful discussions. This work is
supported by CNSF, NSF, and NASA.

% \appendix

% \section{}
% \label{}

% Bibliographic references with the natbib package:
% Parenthetical: \citep{Bai92} produces (Bailyn 1992).
% Textual: \citet{Bai95} produces Bailyn et al. (1995).
% An affix and part of a reference:
%   \citep[e.g.][Ch. 2]{Bar76}
%   produces (e.g. Barnes et al. 1976, Ch. 2).


\begin{thebibliography}{}

% \bibitem[Names(Year)]{label} or \bibitem[Names(Year)Long names]{label}.
% (\harvarditem{Name}{Year}{label} is also supported.)

%\bibitem[]{}
\bibitem{Ni} M. A. Nielsen and I. S. Chuang, \textit{Quantm computation and
quantum information}, (Cambridge University Press, Cambridge, UK,
2000).

\bibitem{Dur} W. D\"{u}r, G. Vidal, and J. I. Cirac, Phys.
Rev. A \textbf{62}, 062314 (2000).

\bibitem{Coffman}V. Coffman, J. Kundu, and W. K. Wootters,
Phys. Rev. A \textbf{61}, 052306 (2000).

\bibitem{Bennett}C. H. Bennett \textit{et al.},
Phys. Rev. A \textbf{63}, 012307 (2001).

\bibitem{Green} D.M. Greenberger \textit{et al.}, Am. J.
Phys. \textbf{58}, 1131(1990).

\bibitem{ut}M. W. Mitchell, J. S. Lundeen, A. M. Steinberg,
Nature {\bf 429}, 161 (2004).

\bibitem{pan}Zhi Zhao, Yu-Ao Chen, An-Ning Zhang, Tao Yang,
Hans J. Briegel, and Jian-Wei Pan, Nature {\bf 430}, 54 (2004).

\bibitem{Molmer} K. Molmer and A. Sorensen, Phys. Rev. Lett.
\textbf{82}, 1835 (1999).

\bibitem{jaksch}D. Jaksch, J. I. Cirac, and P. Zoller,
Phys. Rev. A {\bf 65}, 033625 (2002).

\bibitem{Chat}K. Ch. Chatzisavvas, C. Daskaloyannis, and C.P.Panos,
Quantum Computers and Computing \textbf{4}, 94 (2003).

\bibitem{Sacket}C. A. Sackett {\it et al.}, Nature {\bf 404}, 256 (2000).

\bibitem{You} L. You, Phys. Rev. Lett. \textbf{90}, 030402 (2003).

\bibitem{Got} D. Gottesman, Phys. Rev. A \textbf{54}, 1862 (1996).

\bibitem{Bri} H. J. Briegel and R. Raussendorf, Phys. Rev. Lett. \textbf{86}%
, 910 (2001).

\bibitem{Rau1} R. Raussendorf, D. E. Browne, and H. J. Briegel,
Phys. Rev. A {\bf 68}, 022312 (2003).

\bibitem{Hil} M. Hillery, V. Buzek, and A. Berthiaume,
Phys. Rev. A {\bf 59}, 1829 (1999).

\bibitem{Cle} R. Cleve, D. Gottesman, and H.-K. Lo, Phys. Rev.
Lett. \textbf{83}, 648 (1999).

\bibitem{Ter} B.M. Terhal, D.P. DiVincenzo, and D.W. Leung, Phys.
Rev. Lett. \textbf{86}, 5807 (2001).

\bibitem{Zeng} B. Zeng {\it et al.}, Phys. Rev. A
{\bf 68}, 042316 (2003).

\end{thebibliography}
\end{document}